# SC VALL-E: Style-Controllable Zero-Shot Text to Speech Synthesizer

**DAEGYEOM KIM[1], SEONGHO HONG[1], AND YONG-HOON CHOI[1] (Member, IEEE)**
[1]School of Robotics, Kwangwoon University, Seoul 01897, South Korea

CORRESPONDING AUTHOR: Yong-Hoon Choi (e-mail: yhchoi@kw.ac.kr).

This work was supported in part by the National Research Foundation of Korea (NRF) Grant funded by the Korea Government MSIT under Grant 2021R1F1A1064080.

**ABSTRACT** Expressive speech synthesis models are trained by adding corpora with diverse speakers, various emotions, and different speaking styles to the dataset, in order to control various characteristics of speech and generate the desired voice. In this paper, we propose a style control (SC) VALL-E model based on the neural codec language model (called VALL-E), which follows the structure of the generative pretrained transformer 3 (GPT-3). The proposed SC VALL-E takes input from text sentences and prompt audio and is designed to generate controllable speech by not simply mimicking the characteristics of the prompt audio but by controlling the attributes to produce diverse voices. We identify tokens in the style embedding matrix of the newly designed style network that represent attributes such as emotion, speaking rate, pitch, and voice intensity, and design a model that can control these attributes. To evaluate the performance of SC VALL-E, we conduct comparative experiments with three representative expressive speech synthesis models: global style token (GST) Tacotron2, variational autoencoder (VAE) Tacotron2, and original VALL-E. We measure word error rate (WER), F0 voiced error (FVE), and F0 gross pitch error (F0GPE) as evaluation metrics to assess the accuracy of generated sentences. For comparing the quality of synthesized speech, we measure comparative mean option score (CMOS) and similarity mean option score (SMOS). To evaluate the style control ability of the generated speech, we observe the changes in F0 and mel-spectrogram by modifying the trained tokens. When using prompt audio that is not present in the training data, SC VALL-E generates a variety of expressive sounds and demonstrates competitive performance compared to the existing models. Our implementation, pretrained models, and audio samples are located on GitHub.

## I. INTRODUCTION

A text-to-speech (TTS) system is a software program that converts text into spoken words, aiming to generate natural-sounding voices similar to human speech. With the application of deep learning techniques, significant advancements have been achieved in both the acoustic models [1-6] and vocoders [7-10] used for speech synthesis. In particular, the use of the sequence-to-sequence (seq2seq) [11] model has made it possible to synthesize natural-sounding speech corresponding to text of varying lengths. One notable model is Tacotron [1], which is the first acoustic model based on seq2seq with an attention mechanism. Tacotron2 [2], building upon Tacotron, incorporates the WaveNet vocoder [7] and addresses the attention alignment issue, resulting in more natural speech synthesis compared to traditional statistical parametric methods. However, these models lack the ability to control characteristics such as emotions or speaking styles in the synthesized speech.

The goal of speech synthesis is to generate natural-sounding voices that closely resemble human speech. As a result, research has been conducted to control various aspects such as emotions, intonation, and speaking rate [12-17]. Prosody transfer Tacotron [12] successfully generates synthesized speech for different speakers by employing a reference encoder that combines 2D-convolution and gated recurrent unit (GRU). Global style token (GST) Tacotron [13] extracts style tokens from reference audio that capture the speaker's style, enabling the generation of speech that expresses the speaker's style and emotions. By combining Tacotron 2 with GST, GST Tacotron allows for control over emotions, speaking styles, background noise, and more by manipulating style tokens. Similarly, variational autoencoder (VAE) Tacotron [14] extracts a fixed-length latent vector **z** through VAE from reference audio, thereby generating speech with embedded speaker style and emotions. Li *et al*. [15] demonstrated improved quality of synthesized speech by incorporating two emotion classifiers and four loss functions, thereby allowing control over emotion intensity through the manipulation of style embedding space using values ranging from 1.5 to 2.5.

Several speech synthesis models have been proposed to overcome limitations of Tacotron. Autoregressive models inherently suffer from long inference times, error propagation, and attention alignment issues, leading to less robust performance. To address these issues, non-autoregressive models such as Transformer-TTS [3], FastSpeech [4], FastSpeech2 [5], and FastPitch [6] have been developed. These models improve inference speed and exhibit a more robust structure with reduced skipping and repeating issues. Du *et al*. [16] enhance the performance of FastSpeech2 by introducing a Gaussian mixture model (GMM)-based prosody predictor and extracting phoneme-level prosody from reference audios. This approach results in more natural-sounding speech synthesis. By training 20 GMMs for phonetic prosody representation and incorporating sampling during inference, speech with diverse prosody can be generated. This research demonstrates that phoneme-level embeddings accurately capture prosody compared to utterance-level embeddings. Guo *et al*. [17] utilize decision trees to classify prosody representations based on word pronunciation, followed by GMM clustering to generate prosody tags. Training prosody tags for each word enables not only more natural speech synthesis but also the ability to control word-level prosody by manipulating the prosody tags. This approach, built upon FastSpeech2 and employing a tree structure, offers the advantage of easy application to other language speech synthesis problems. Moon *et al*. [18] conducted a study on extracting speaker characteristics by proposing a reference encoder that applies image style transfer techniques to mel-spectrogram conversion. This approach explores the use of image style transfer methods to capture speaker characteristics in the synthesis process.

Recent advancements in speech synthesis have relied on clean corpora recorded in studios by voice actors to achieve high-quality synthesized speech. However, using large-scale datasets collected from the Internet often fails to meet these requirements, leading to performance degradation. Consequently, researchers have resorted to using high-quality but limited training data, leading to the challenge of reduced generalization capability when synthesizing speech for speakers not included in the training data. To tackle these challenges, previous studies have explored speaker adaptation [19] and speaker encoding [20], [21] techniques. However, speaker adaptation methods tend to exhibit a significant decrease in synthesis quality when there is insufficient or low-quality voice data available for the target speaker. Additionally, adapting the model for new speakers may result in lower synthesis quality compared to existing speakers. Speaker encoding methods encounter difficulties in capturing detailed characteristics such as the speaker's speech state, emotions, and intonation. Moreover, these methods struggle to adequately encode speakers not included in the training data, leading to diminished generalization capability.

Motivated by recent advancements in natural language generation models, ongoing studies have focused on language models trained on large and diverse multi-speaker speech datasets to improve generalization performance [22-26]. PromptTTS [22] generates speech by utilizing prompt texts that provide descriptions of the desired speech. It incorporates style encoder, content encoder, and speech decoder to generate speech based on the provided prompts. VALL-E [23] has the capability to synthesize high-quality personalized speech using 3-second recordings as prompts. VALL-E leverages EnCodec [27], a neural network-based audio codec model, to incorporate speech waveforms into the language model. EnCodec employs an encoder and decoder structure based on residual vector quantization (RVQ) [28] to compress speech waveforms into discrete tokens that contain speaker and acoustic information. These tokens are then restored into high-quality waveforms. Through these approaches, it has been observed that speech synthesis can preserve the speaker's emotions and acoustic environments as specified in the prompts.

In this paper, we propose Style Control (SC) VALL-E, an extension of the VALL-E model that allows for the control of synthesized speech's emotions, speaker styles, and various acoustic features. SC VALL-E is an end-to-end zero-shot TTS model that incorporates style tokens. During the training process of SC VALL-E, the style tokens learn the speech's style and features from the output of the acoustic prompt encoder. During speech synthesis, we can control the style and features associated with each token by multiplying them with values ranging from 0.5 to 2.5, enabling the generation of desired speech.

To validate the proposed model, we conducted objective and subjective evaluations using voice prompts from speakers not included in the training data. Objective evaluations included metrics such as F0 voiced error (FVE) [29], F0 gross pitch error (F0GPE) [30], and word error rate (WER) based on the audio. WER was calculated by converting the synthesized speech into text using OpenAI Whisper model and comparing it with the input text to assess the model's faithfulness in synthesizing speech corresponding to the input text. Subjective evaluations were conducted using comparative mean option score (CMOS) and similarity mean option score (SMOS). Additionally, we visualized and compared the mel-spectrogram and F0 values of the synthesized speech to observe changes when controlling the style. The experimental results confirmed that the proposed model generates expressive speech, even for speakers not represented in the training data. It was observed that controlling the style tokens allowed for manipulation of acoustic features such as speech rate, pitch, volume, and emotions. Through the experiments, we confirmed that the synthesized speech closely resembled the prompt audio, and the SMOS values were higher compared to the comparative models, indicating a higher level of similarity.

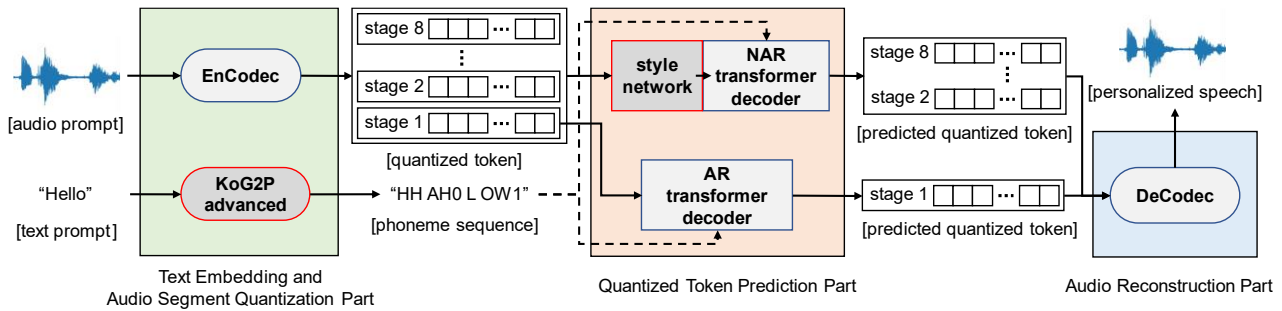

**FIGURE 1.** SC VALL-E architecture. The added blocks to the original VALL-E model [23] are indicated in red color.

The rest of this paper is organized as follows. In Section II, the proposed model is described. Sections III and IV explain the experimental setup and present the experimental results of the proposed model. Finally, in Section V, the conclusions of the study are provided along with a discussion on future research directions.

## II. MODEL DEVELOPMENT

The overall architecture of the proposed SC VALL-E in this paper consists of three main components: the Text Embedding and Audio Segment Quantization part, the Quantized Token Prediction part, and the Audio Reconstruction part, as depicted in Figure 1. The original VALL-E architecture [23] serves as the underlying base model for generating high-quality synthesized speech in zero-shot scenarios.

### A. SC VALL-E ARCHITECTURE

The objective of SC VALL-E is to train the style network, allowing for the control of various voice characteristics such as emotion, speaker's speaking style, speech rate, intonation, and prosody. SC VALL-E comprises text embedding and audio quantization modules designed for Korean speech synthesis. Utilizing each Korean pronunciation symbol as an individual token would result in an excessively large number of tokens, making training inefficient. Additionally, the sounds can vary depending on whether symbols are in the initial or final positions, even for the same Korean alphabet. To address this, tokenization is performed by separating the symbols into initial consonants, vowels, and final consonants. These tokens serve as inputs for the Text Embedding layer, and the output is then used as inputs for both the autoregressive (AR) and non-autoregressive (NAR) blocks in the Quantized Token Prediction component. To convert prompt audio into quantized tokens, a pretrained encoder model is employed, with an encoder bandwidth of 6 Kbps used for acoustic token extraction. The quantized tokens follow a residual vector quantization (RVQ) structure comprising a total of 8 layers. The first layer serves as input for the AR model, while the remaining 7 layers are used as inputs for the NAR model.

The Quantized Token Prediction part is responsible for predicting the quantized tokens required to synthesize personalized speech that corresponds to the given text. It utilizes both the text embedding and the quantized tokens. This prediction part consists of two blocks: the autoregressive (AR) block and the non-autoregressive (NAR) block, which follow a transformer decoder architecture. The AR block employs a causal attention mechanism, considering only the previous quantized tokens to calculate attention scores. Since the initial stage (i.e., stage 1) of the quantized tokens contains acoustic characteristics such as speaker information, this approach performs well in predicting speech length that aligns with the given text for various speakers. On the other hand, the NAR block adopts a non-causal attention mechanism, considering the entire set of quantized tokens to calculate attention scores. The subsequent stages (i.e., stage 2 to 8) of the quantized tokens contain detailed acoustic information that significantly influences the quality of synthesized speech. The NAR block incorporates the proposed style network, enabling the style tokens to learn prominent features from the acoustic details. This allows the tokens to capture various aspects, including speaking rate, speech volume, intensity of emotions, intonation, pitch, and background noise. By changing the value of tokens, control and scaling of acoustic features become feasible.

The Audio Reconstruction part utilizes a pretrained DeCodec [27] block to convert the predicted quantized tokens from the Quantized Token Prediction part back into audio waveforms. During this process, the AR model predicts the temporal length of the quantized tokens, determining the overall size of the token sequence. Subsequently, the NAR model predicts the values of the remaining tokens and generates personalized speech tokens. By leveraging the pretrained DeCodec block, the quantized tokens are transformed back into reconstructed audio waveforms. In summary, the AR model's role is to predict the temporal length of the quantized tokens, while the NAR model is responsible

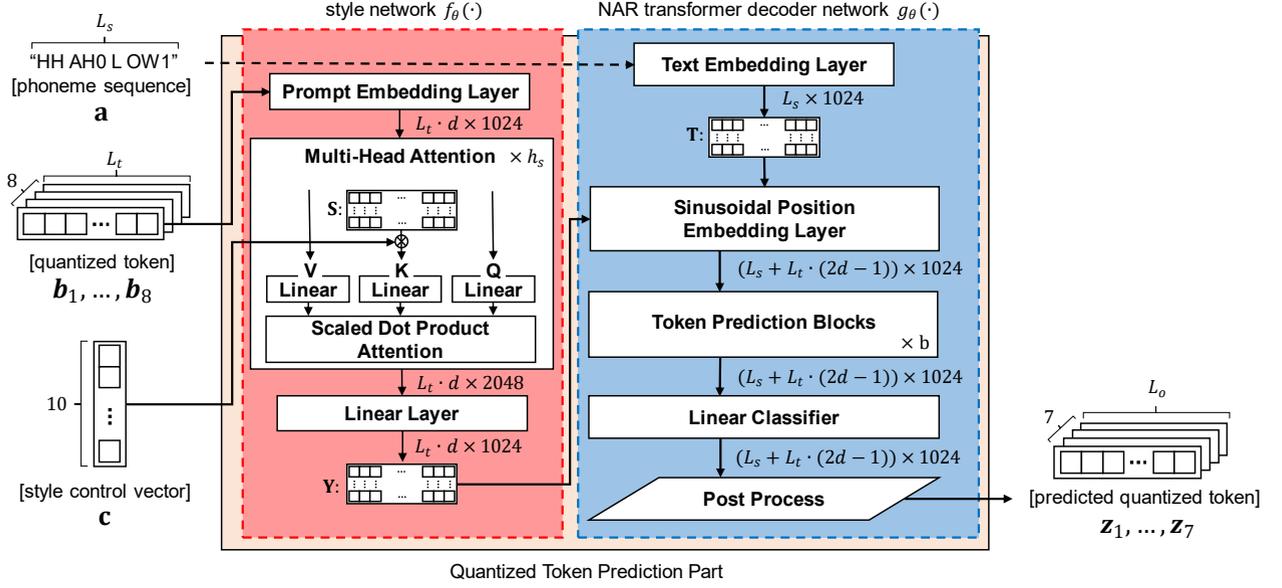

**FIGURE 2.** Quantized Token Prediction Part of SC VALL-E. The style network on the left enables control over the style of the synthesized speech.

for predicting the values of the remaining tokens and generating personalized speech tokens.

### B. THE QUANTIZED TOKEN PREDICTION PART

The goal of the Quantized Token Prediction part is to learn eight representations, $\mathbf{z}_1, \ldots, \mathbf{z}_8$, which are referred to as predicted quantized tokens. As shown in Figure 2, the NAR transformer decoder utilizes the style network to learn prominent features of the audio from quantized tokens. The AR transformer decoder, not depicted in Figure 2, has the same structure and hyperparameters as the NAR transformer decoder, except for the attention mechanism.

Let vectors $\mathbf{b}_i \in \mathbb{R}^{L_t}$ be quantized tokens with the same length $L_t$, where $i \in \{1, \ldots, 8\}$. $L_T$ is determined by the length of the audio prompt. Quantized tokens $\mathbf{b}_1, \ldots, \mathbf{b}_8$ are sequentially inputted into the Prompt Embedding layer, resulting in the output of the style token $\mathbf{S} \in \mathbb{R}^{N \times 256}$, where $N$ represents the number of 256-length style tokens set to $N = 10$. $\mathbf{S}$ is then used as an input to the 8-head attention, where attention scores are computed. It extracts style within $\mathbf{S}$ and captures local variations in the acoustics. If $N$ is too large, the prominent acoustic features of each token may not be learned. Conversely, if $N$ is too small, there is a risk of missing out on the opportunity to learn the diverse acoustic characteristics. We use the value of $N = 10$, which we experimentally confirmed to be the best option.

Let vector $\mathbf{c} \in \mathbb{R}^N$ be a style control vector of length $N$, where $0.5 \leq c_i \leq 2.5$ for $i \in \{1, \ldots, N\}$. $N$ represents the number of style tokens, and it is set to 10. It has been confirmed that $c_i$ should be positive, and setting it to a value greater than 2.5 leads to excessive transformation of the attention key, resulting in the synthesis of incomprehensible sounds or the problem of repetitively synthesizing certain words. The column vector $\mathbf{c}$ and the row dimension of matrix $\mathbf{S}$ both have a size of $N$, and by broadcasting $\mathbf{c}$ to all columns of $\mathbf{S}$ and performing an element-wise product, we obtain $\mathbf{K} \in \mathbb{R}^{N \times 256}$. $\mathbf{K}$ is used as the key in multi-head attention to control the acoustic features. Through experiments, it has been confirmed that using $\mathbf{K}$ as the key in multi-head attention allows for more consistent control compared to when it is used as the query or value. During training, $\forall i \in \{1, \ldots, N\}: c_i = 1$, while during inference, $\forall i \in \{1, \ldots, N\}: 0.5 \leq c_i \leq 2.5$ is assigned to express various styles.

Let's denote $\mathbf{Y} \in \mathbb{R}^{L_t \cdot d \times 1024}$ as the style embedding matrix, where integer $d \in \{2, \ldots, 8\}$ is a value randomly determined for each sequence in the Prompt Embedding layer. In summary, as shown in Figure 2, intermediate representation $\mathbf{Y}$ can be obtained by passing through the attention block and linear layer. In summary, $\mathbf{Y}$ is obtained by style function $f_\theta(\cdot)$ parameterized with $\theta$:

$$\mathbf{Y} = f_\theta(\mathbf{b}_1, \ldots, \mathbf{b}_8, \mathbf{c}). \quad (1)$$

Let vector $\mathbf{a} \in \mathbb{R}^{L_s}$ be phoneme sequence with length $L_S$. In the case of the AR model, the input to the transformer decoder consists of $\mathbf{a}$ and $\mathbf{b}_1$. In the case of the NAR model, the input consists of $\mathbf{a}$ and $\mathbf{Y}$. Phoneme sequence $\mathbf{a}$ passes through the Text Embedding layer to be embedded into the text embedding matrix $\mathbf{T} \in \mathbb{R}^{L_s \times 1024}$. $\mathbf{T}$ is then inputted into the Sinusoidal Position Embedding layer along with $\mathbf{Y}$ and additional acoustic information from $d-1$ stage is incorporated. Through this process, the model learns the positional information of $\mathbf{T}$ and the temporal characteristics of $\mathbf{Y}$, enabling it to learn the connections between each stage and generate high-quality speech, facilitating the generation of high-quality speech. The Token Prediction block and Linear Classifier block are used to predict acoustic tokens, and the Post Processing block performs a resize operation to align the temporal length of the acoustic token vectors with the text.

Let vectors $\mathbf{z}_i \in \mathbb{R}^{L_O}$ be predicted quantized tokens with the same length $L_O$, where $i \in \{1, \dots, 7\}$. $L_O$ is determined by the length of the input sequence. Outputs $\mathbf{z}_1, \dots, \mathbf{z}_7$ are obtained by decoding function $g_\theta(\cdot)$ parameterized with $\theta$:

$$\mathbf{Z} = C_h(\mathbf{z}_1, \dots, \mathbf{z}_7) = g_\theta(\mathbf{a}, \mathbf{Y}, \mathbf{b}_1, \dots, \mathbf{b}_8), \quad (2)$$

where $C_h(\cdot)$ is a function that concatenates vectors to create a single-row matrix.

The Token Prediction block in NAR model is composed of multi-head attention and a feed forward network. Adaptive normalization [32] is employed to adaptively adjust the mean and standard deviation based on the characteristics of the input data, enabling easier adaptation to the diversity of the input data and enhancing the flexibility and expressiveness of the model. Moreover, it facilitates smooth information propagation between layers and ensures smooth gradient flow, preventing significant gradient fluctuations during early training stages and enabling stable learning. The input to the block captures important aspects of the local text and acoustic representations through multi-head attention with 16 heads, learning the temporal information of the acoustic representation required to generate natural-sounding speech.

### III. MODEL TRAINING AND EVALUATION

To validate the performance of SC VALL-E, three types of experiments are conducted. The first experiment involves comparing the synthesized voices of the VALL-E [23] model with those of the proposed model to determine if the proposed model can synthesize more expressive voices. In the second experiment, GST-Tacotron [13] and VAE-Tacotron [14] are compared to assess their ability to synthesize voices that accurately represent the style of speakers not present in the training data when using 3-second, 5-second, and 7-second voice prompts. The third experiment aims to examine whether the acoustic characteristics of the synthesized voices can be controlled by manipulating the style tokens. This is achieved by visualizing the mel-spectrogram and F0 of the generated voices. Our implementation, pretrained models, and audio samples can be found on GitHub [33].

#### A. DATASETS

The speech datasets used in this experiment was collected from artificial intelligence hub (AI Hub), a platform operated by the national IT industry promotion agency (NIPA) that provides open AI-related datasets. The collected datasets are categorized into five main categories: command speech dataset, conversation speech dataset, emotion and speaking style-specific speech dataset, Korean speakers' foreign language speech dataset, and Korean speech dataset.

The command speech dataset and conversation speech dataset are further divided into three categories: male/female, infants/children, and elderly male/female. The collected corpora include speeches recorded by professional voice actors in a studio, as well as speeches collected by individuals using personal recording devices in both indoor and outdoor settings, representing a wide range of quality levels. More detailed information about the collected corpora can be found in Table 1. Most of the audio files have an average duration of 3 to 5 seconds, with the characteristic of elderly male/female speakers generally having longer utterance durations. The datasets include a total of 14,050 speakers, and the total duration of the audio files is 21,497 hours.

**TABLE 1.** Datasets configuration. Dataset details can be accessed on https://github.com/0913ktg/0913ktg.github.io.

| Corpus | Number of speakers | Number of audio clips (hours) | Average duration |
|---|---|---|---|
| Command speech (male/female) | 2,000 | 3,550,288 (2,885 h) | 2.9 s |
| Command speech (infants/children) | 2,000 | 2,943,946 (2,415 h) | 2.9 s |
| Command speech (elderly male/female) | 1,000 | 2,475,227 (2,449 h) | 3.5 s |
| Conversation speech (male/female) | 2,000 | 2,522,014 (3,538 h) | 5.1 s |
| Conversation speech (infants/children) | 2,000 | 2,848,385 (2,735 h) | 3.5 s |
| Conversation speech (elderly male/female) | 1,000 | 1,277,524 (2,717 h) | 7.6 s |
| Emotion and speaking style-specific speech | 50 | 927,635 (953 h) | 3.7 s |
| Korean speakers' foreign language pronunciation speech | 2,000 | 2,412,259 (3,101 h) | 4.4 s |
| Korean speech dataset | 2,000 | 565,356 (704 h) | 4.4 s |

The command speech dataset consists of simple sentence utterances spoken by individuals in their daily life to an AI assistant. The conversation speech dataset consists of simple sentence utterances spoken by individuals in their daily life. The emotion and speaking style-specific speech dataset consists of speeches recorded by 50 professional voice actors in a studio, representing 7 emotions and 5 speaking styles. The Korean speakers' foreign language speech dataset consists of simple sentence utterances containing foreign words spoken by individuals. The Korean speech dataset consists of conversational speech recorded by two individuals on various topics, segmented into sentence units.

The training and validation data are split in an 8:2 ratio, and the sampling rate for all the data used in training and validation is set to 24,000 Hz. The test data consist of speeches from unseen speakers, which is either directly recorded or collected from the Internet.

#### B. MODEL SETUP

The proposed SC VALL-E implementation utilized the PyTorch framework and the DeepSpeed optimization library. The dimensions for both the text embedding and prompt embedding are set to 1024. The number of heads in the style token part is set to 8, the embedding dimension of the style network is set to 2048, and the number of style tokens is set to 10, The network consists of 12 Token Prediction blocks, with a token dimension of 256. The attention dimension within each block is set to 512, the hidden layer dimension of the feed-forward network is set to 2048, and the output dimension is set to 512.

**TABLE 2.** Considered hyperparameters for SC VALL-E.

| Layer | Hyperparameters | Values |
|---|---|---|
| Style network | Length of prompt embedding | 1024 |
| | Number of head ($h_s$) | 8 |
| | Number of style tokens ($N$) | 10 |
| Transformer decoder network | Length of text embedding | 1024 |
| | Number of Token Prediction Blocks ($b$) | 12 |
| Training setup | Learning rate | 0.0002 |
| | Batch size | 20 |
| | Sampling temperature | 0.2 |
| | Gradient skipping | 100 |

The optimizer used is Adam, with a learning rate of 0.0002. The batch size is set to 20, and the weight decay linearly decreased from 0.0002 to $1 \times 10^{-6}$ every 1000 steps. The sampling temperature is set to 0.2, and gradient skipping is set to 100. The tensor type used is float16. Sampling temperature is a hyperparameter that controls the randomness of the sampling process. The main hyperparameters used in SC VALL-E are summarized in Table 2.

The workstation used for training has the Ubuntu 20.04 LTS operating system, and software dependencies are managed using Nvidia Docker. The training of the proposed SC VALL-E model and the original VALL-E model utilizes four Nvidia A100 GPUs with 80 GB of memory each. For the training of GST-Tacotron, VAE-Tacotron, and HiFi-GAN [10], one NVIDIA A100 GPU with 80 GB of memory is used for each model.

### C. TRAINING AND INFERENCE

To accelerate the training time, the script is converted into a phoneme sequence using KoG2Padvanced [31] and saved. The audio files are then converted into quantized tokens using a pretrained EnCodec [27] model and saved as well. Both the SC VALL-E and the original VALL-E models are trained using the prepared phoneme sequences and quantized tokens.

For inference, the relevant parts of the model are extracted from the DeepSpeed checkpoint, and speaker information is added and saved separately. Speech synthesis is performed using the AR and NAR model checkpoints.

To evaluate the performance of the models, objective evaluation, subjective evaluation, and style control visualization are performed. Objective evaluation is done using WER, FVE, and F0GPE. WER measures the error rate between the recognized text and the reference text. We use OpenAI's Whisper to transcribe the synthesized speech into text and calculate the minimum number of word-level edit operations required to transform the recognized text into the reference text. WER is calculated by

$$WER = \frac{S + D + I}{N},\quad (3)$$

where, $S$ is the number of substitution errors (words replaced with different words), $D$ is the number of deletion errors (words missing from the recognized output), $I$ is the number of insertion errors (extra words inserted into the recognized output), and $N$ is the total number of words in the reference text. WER represents the ratio of word-level errors between the output of the given system and the reference text. A lower WER indicates a higher system performance.

FVE and F0GPE are indicators representing the quality of voice using fundamental frequency F0, and the unit of both indicators is [%]. FVE is given by

$$\text{FVE}(f, \hat{f}) = \sqrt{\frac{\sum_{i=1}^{N}(f_i - \hat{f}_i)^2}{N}},\quad (4)$$

where $f$ and $\hat{f}$ are F0 values of original and synthesized voice, respectively, and $N$ is the total number of F0's in the original speech. F0GPE is given by

$$\text{F0GPE} = \frac{N_{F_0 E}}{N_{VV}} \times 100,\quad (5)$$

where $N_{F_0 E}$ is the number of voiced frames when the estimated F0 values are incorrect, and $N_{VV}$ is the total number of voiced frames. The incorrect F0 value is defined as a value that falls outside $\delta$ (typically $\pm 20\%$) of the correct F0 value. The smaller the value of all the evaluation indicators, the better the performance.

For subjective evaluation, 30 Korean participants evaluate 120 audio clips. The evaluation is conducted using the CMOS and SMOS. The CMOS ranges from -3 (i.e., comparative system is significantly worse than the reference) to 3 (i.e., comparative system is significantly better than the reference). The SMOS indicates how well the synthesized speech reflects the speaker style of the original speech on a scale of 1 to 5. The audio clips used for subjective evaluation can be found in https://0913ktg.github.io/.

### IV. RESULT

The proposed model was not sufficiently trained due to the utilization of a large-scale training dataset. The total size of the dataset used for training and validation amounts to 6.3 TB, which corresponds to a substantial volume equivalent to 21,497 hours. The proposed model was trained for only 1.5 epochs over a period of 68 days using four NVIDIA A100 GPUs with 80 GB of memory. In contrast, the comparison models such as GST-Tacotron and VAE-Tacotron were trained to a satisfactory extent.

### A. OBJECTIVE PERFORMANCE EVALUATION

The results of the objective performance evaluation comparing the proposed method with the existing methods are presented in Table 3. GST-Tacotron and VAE-Tacotron are models that take only text and reference audio as inputs. VALL-E is a model that utilizes text and reference audio as inputs, incorporating KoG2Padvanced. SC VALL-E is the model proposed in this paper, which incorporates KoG2Padvanced and our style network.

**TABLE 3.** Objective performance evaluation of speech synthesis using unseen speaker's audio input. Values in ( ) are 95% confidence intervals.

| Model | WER | FVE | F0GPE |
|---|---|---|---|
| GST-Tacotron [13] | **0.21** (±0.05) | 64.24 (±6.25) | 54.53 (±4.50) |
| VAE-Tacotron [14] | 0.23 (±0.05) | **54.75** (±5.12) | **44.79** (±3.72) |
| Original VALL-E [23] | 0.30 (±0.05) | 64.75 (±5.12) | 56.79 (±3.72) |
| SC VALL-E (ours) | 0.25 (±0.05) | 64.19 (±6.25) | 56.53 (±4.50) |

**TABLE 4.** Subjective performance evaluation of speech synthesis using unseen speaker's audio input. CMOS and SMOS evaluation. The CMOS score is the result of comparison with the proposed SC VALL-E model. Values in ( ) are 95% confidence intervals.

| Model | CMOS | SMOS |
|---|---|---|
| GST-Tacotron [13] | −1.12 (±0.10) | 3.70 (±0.12) |
| VAE-Tacotron [14] | −0.92 (±0.08) | 3.90 (±0.09) |
| Original VALL-E [23] | −0.40 (±0.09) | 3.90 (±0.10) |
| SC VALL-E (ours) | - | **4.10** (±0.10) |

From Table 3, it can be observed that the proposed model, SC VALL-E, achieves slightly lower evaluation scores compared to GST-Tacotron and VAE-Tacotron, while showing similar scores to the original VALL-E model. The lower scores of the VALL-E-based model can be attributed to its design of generating diverse voices by leveraging the style of the reference audio, rather than generating data similar to the training data.

WER is calculated by converting the synthesized speech into text using OpenAI Whisper model and comparing it with the input text to assess how faithfully the model synthesizes speech corresponding to the input text. The reason for the higher WER of the proposed model compared to the Tacotron-based model is that the proposed model lacked sufficient training for specific words or sentences. As a result, some words were skipped during audio generation, and the model failed to synthesize speech that was recognizable by the Whisper model.

It was observed that the high F0GPE is due to changes in the synthesized speech style, resulting in differences in speaker-specific characteristics such as breathing sounds and pronunciation. The high FVE can be attributed to variations in speech rate, timing, and word-specific speech length in the synthesized speech.

Upon examining the synthesized speech using reference audios of 3 seconds, 5 seconds, and 7 seconds, it was observed that SC VALL-E captured the speaking style of the reference speaker the best, especially with the shorter 3-second reference audio. In contrast, GST-Tacotron and VAE-Tacotron failed to accurately replicate the speaker style of the reference audio when using shorter reference audios. Instead, they synthesized speech using the style of other speakers present in the training data. Comparing it to the original VALL-E, the proposed model demonstrated the ability to effectively represent the speaker's style regardless of the length of the reference audio.

### B. SUBJECTIVE PERFORMANCE EVALUATION

Table 4 presents the evaluation results for CMOS and SMOS, comparing the proposed model with the comparison models. SMOS evaluates the resemblance of the synthesized speech to the speaker's speaking style, while CMOS compares the performance of the proposed model and the comparison models in capturing the speaker's speaking style. Both evaluations primarily focus on assessing how well the models replicate the speaker's speaking style, rather than the overall quality of the synthesized speech.

The CMOS and SMOS results indicate that the proposed model outperforms the comparison models in reflecting the speaking style of the given speaker in a zero-shot setting. This can be attributed to the multi-head attention of the style token, which effectively captures both major characteristics such as speech rate, voice size, and pitch, as well as finer details like the speaker's speech habits, resonances, and tremors. As a result, the proposed model exhibits superior performance compared to the original VALL-E model. In contrast, the synthesized speech produced by GST-Tacotron and VAE-Tacotron does not resemble the speaker in the reference audio, as it is synthesized using the voices of speakers from the training data.

### C. STYLE CONTROL VISUALIZATION

Figure 3 illustrates the visualization of the mel-spectrogram and F0 of the synthesized speech when controlling the audio style. As mentioned earlier, to control style, the style control vector

$$\mathbf{c} = \begin{bmatrix} c_1 \\ \vdots \\ c_{10} \end{bmatrix}, 0.5 \leq c_i \leq 2.5 \quad (6)$$

is broadcasted to all columns of style token matrix $\mathbf{S} \in \mathbb{R}^{10 \times 256}$ and is performed an element-wise product. In this experiment, the prompt audio consisted of a Korean sentence, "An-nyeong-ha-se-yo. Ban-gab-seub-ni-da." (Hello, nice to meet you.), spoken in a neutral tone by a female speaker not present in the training data. First, we controlled the audio style

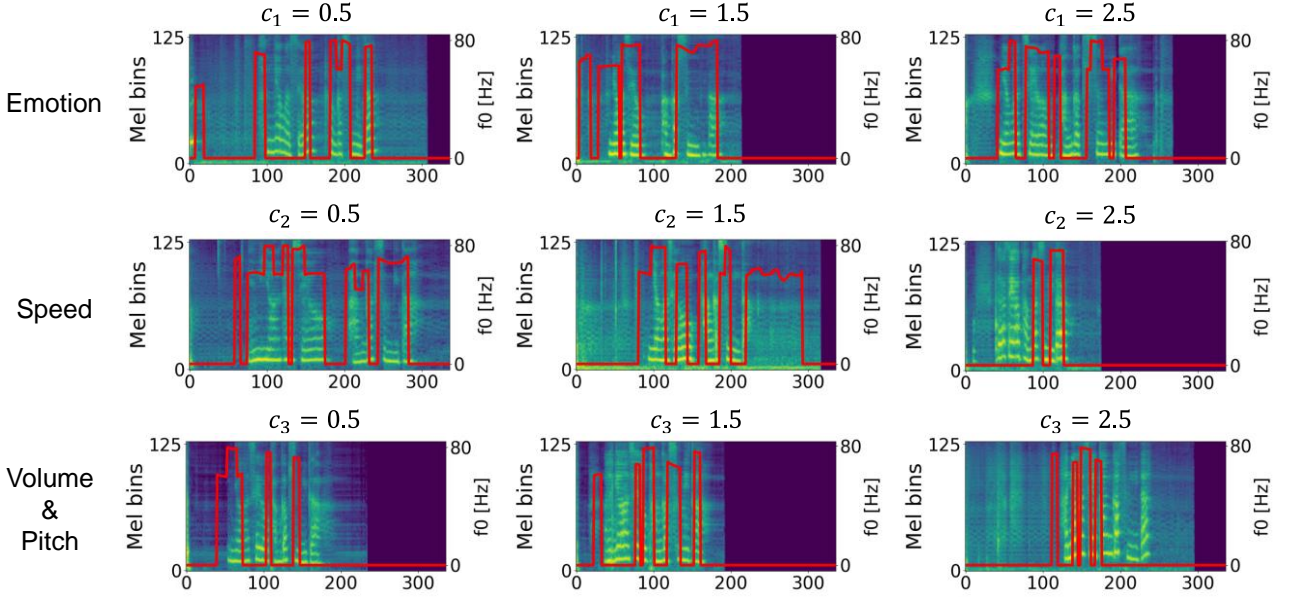

**FIGURE 3.** Changes in mel-spectrogram and F0 when style token control is performed. The top represents the results of controlling the emotion of the generated speech by varying the value of $c_1$ in the style control vector. The middle shows the results of controlling the speed of the generated speech by varying the value of $c_2$. The bottom presents the observations of controlling the volume and pitch of the generated speech by varying the value of $c_3$.

by performing broadcast and multiply operations with **S** using the following three style control vectors:

$$\mathbf{c}^{(1)} = \begin{bmatrix} 0.5 \\ 1 \\ \vdots \\ 1 \end{bmatrix}, \mathbf{c}^{(2)} = \begin{bmatrix} 1.5 \\ 1 \\ \vdots \\ 1 \end{bmatrix}, \mathbf{c}^{(3)} = \begin{bmatrix} 2.5 \\ 1 \\ \vdots \\ 1 \end{bmatrix}. \quad (7)$$

When using style control vectors $\mathbf{c}^{(1)}$, $\mathbf{c}^{(2)}$, and $\mathbf{c}^{(3)}$, it was observed that the emotions transformed into happiness, anger, and sadness, respectively. This can be attributed to the fact that the first row of **S** learned 'emotions' from quantized tokens, allowing effective control of distinguishable emotions such as happiness, anger, and sadness through $c_1$. The changes in mel-spectrograms and F0 according to emotional variations are depicted at the top of Figure 3.

Secondly, we controlled the audio style by performing broadcast and multiply operations with **S** using the following three style control vectors:

$$\mathbf{c}^{(4)} = \begin{bmatrix} 1 \\ 0.5 \\ 1 \\ \vdots \\ 1 \end{bmatrix}, \mathbf{c}^{(5)} = \begin{bmatrix} 1 \\ 1.5 \\ 1 \\ \vdots \\ 1 \end{bmatrix}, \mathbf{c}^{(6)} = \begin{bmatrix} 1 \\ 2.5 \\ 1 \\ \vdots \\ 1 \end{bmatrix}. \quad (8)$$

When using style control vectors $\mathbf{c}^{(4)}$, $\mathbf{c}^{(5)}$, and $\mathbf{c}^{(6)}$, it was observed that the speech rate increased. As larger values are used for $c_2$, the generated speech becomes faster. This can be attributed to the fact that the second row of **S** learned 'tempo' information of the audio from quantized tokens, enabling control of speech rate through $c_2$. The changes in mel-spectrogram and F0 corresponding to variations in speech rate are depicted in the middle section of Figure 3.

Lastly, we controlled the audio style by performing broadcast and multiply operations with **S** using the following three style control vectors:

$$\mathbf{c}^{(7)} = \begin{bmatrix} 1 \\ 1 \\ 0.5 \\ 1 \\ \vdots \\ 1 \end{bmatrix}, \mathbf{c}^{(8)} = \begin{bmatrix} 1 \\ 1 \\ 1.5 \\ 1 \\ \vdots \\ 1 \end{bmatrix}, \mathbf{c}^{(9)} = \begin{bmatrix} 1 \\ 1 \\ 2.5 \\ 1 \\ \vdots \\ 1 \end{bmatrix}. \quad (9)$$

When using style control vectors $\mathbf{c}^{(7)}$, $\mathbf{c}^{(8)}$, and $\mathbf{c}^{(9)}$, an increase in the volume and pitch of the audio was observed. This can be attributed to the fact that the third row of **S** learned 'amplitude and pitch' information of the audio from quantized tokens, allowing control of the audio volume and pitch through $c_3$. The changes in mel-spectrogram and F0 corresponding to variations in volume and pitch are depicted at the bottom of Figure 3.

In addition to the audio prompt and sentence shown in Figure 3, various experiments were conducted with different sentences and audio prompts from diverse speakers to generate different voices. When controlling tokens in **S** with style control vectors $\mathbf{c}^{(1)}$ through $\mathbf{c}^{(9)}$ consistent variations were observed. Overall, these results demonstrate that the proposed model effectively controls various aspects of speech synthesis, including emotions, speech rate, and speech characteristics such as volume and pitch, by transforming the information encoded in the style tokens.

We conducted experiments by varying the values of $c_4$ to $c_{10}$, but we did not observe any prominent changes in the generated sounds. Therefore, it is expected that the remaining tokens in **S** did not learn distinctive features. Although

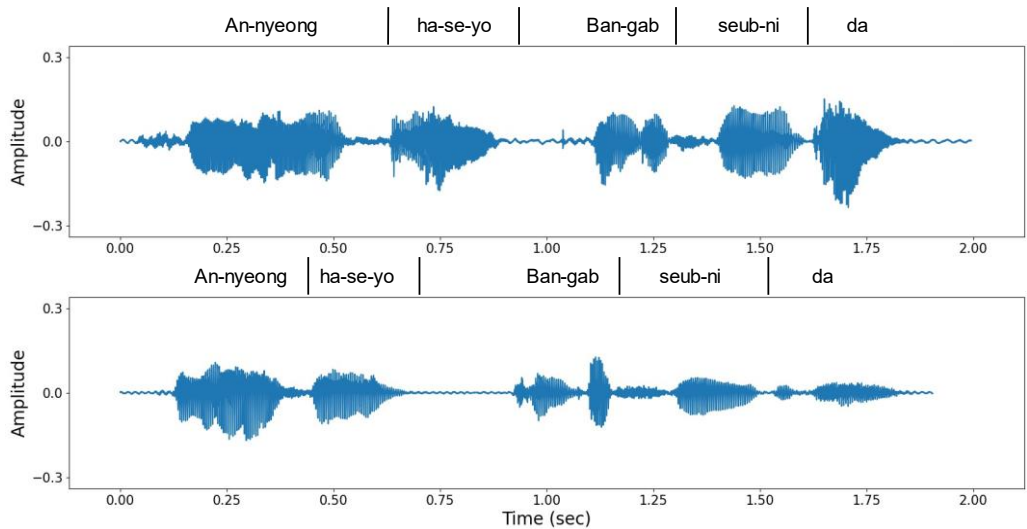

(a) Changes in audio waveform based on the value of $c_1$. Top: $c_1 = 0.5$, Bottom: $c_1 = 1.5$.

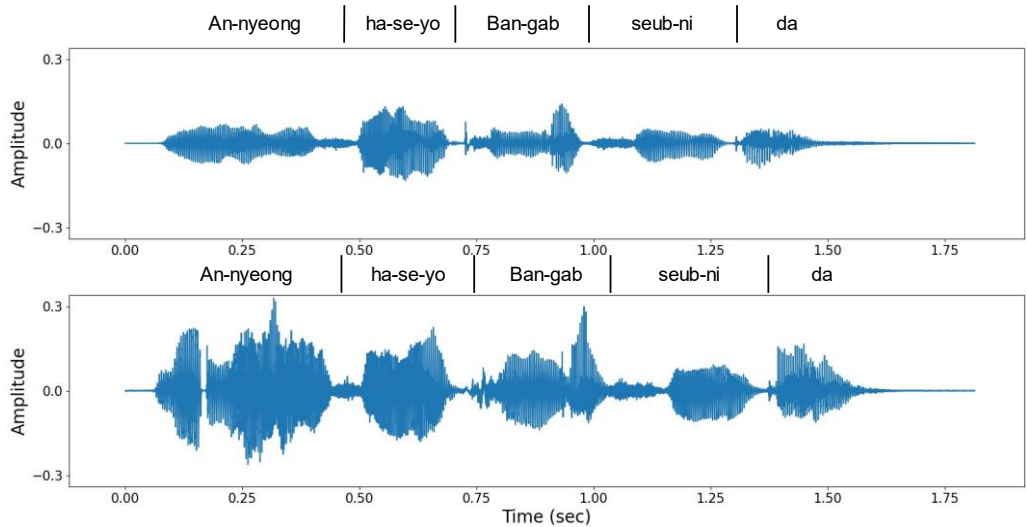

(b) Changes in audio waveform based on the value of $c_3$. Top: $c_3 = 0.5$, Bottom: $c_3 = 1.5$.

**FIGURE 4.** The generated audio waveforms. Input text is "An-nyeong-ha-se-yo. Ban-gab-seub-ni-da."

training allows style tokens to learn additional acoustic characteristics, due to the significantly large scale of the dataset, currently not all tokens have adequately learned the acoustic features. Training is still ongoing, and the performance evaluation results will be updated and made publicly available on our GitHub [33].

### D. SYNTHETIC SPEECH WAVEFORM ANALYSIS

While previous TTS systems exhibit a strong one-to-one mapping between input text and output waveform, VALL-E produces diverse outputs for the same input text due to the randomness in the sampling-based inference process. However, VALL-E lacks explicit control over the desired style output as it does not have a style control input. To address this issue, SC VALL-E introduces a style control vector **c** as an additional input, enabling control over the output. In the experiment, the same input text is used, and the resulting audio waveforms are observed and visualized by varying the values of $c_1$ and $c_3$ from 0.5 to 1.5.

In Figure 4(a), it is evident that the overall shapes of the two audio waveforms are different, as well as the speech onset times and durations of each word. The variation in speech onset times and durations allows the voices to have different styles. The second voice gradually decreases the amplitude of the waveform towards the end, creating a sense of sadness in its expression. In Figure 4(b), the overall shapes of the two waveforms and the speech onset times and durations of each word are similar, but it is clear that the amplitude of the second voice is generally higher. Therefore, it is possible to maintain similar acoustic characteristics while controlling the volume.

Although not depicted in Figure 4(b), pitch variations can also be detected when listening to the voices. Refer to the F0 variations represented at the bottom of Figure 3 for the pitch variations.

## IV. CONCLUSIONS AND FUTURE WORKS

In this paper, we propose the SC VALL-E model, which integrates a style network into the original VALL-E model. We utilize a Korean grapheme-to-phoneme converter to extract phonemes from Korean text and use them for training. To control various features, including the speaking style of the synthesized speech, we multiply values ranging from 0.5 to 2.5 to the style tokens during inference, achieving expressive and dynamic speech generation. While the proposed model exhibits lower evaluation scores compared to GST-Tacotron and VAE-Tacotron in terms of WER, FVE, and F0GPE, it achieves the highest scores in CMOS and SMOS evaluations. In style control experiments, we confirm that the style tokens learn features of emotion, speaking rate, speech intensity, and pitch characteristics. By utilizing the proposed model, it is possible to generate voices that reflect the characteristics of the speakers and enable speech synthesis with richer and more diverse expressions. Therefore, the proposed SC VALL-E model, as a zero-shot learning model, can be utilized in audio content production that requires diverse speakers and varying speaking styles without the need for personalized data collection. We believe that using the proposed model can provide more engaging services in audio content production where the characteristics of synthesized speech need to be frequently changed.

We are continuing our research on improving the training of the style network to make it more stable and efficient. To enhance training efficiency, we are investigating optimization methods that utilize the Audio BucketIterator to create batches using audio samples of the same length and dynamically adjust the batch size based on the length of the audio in order to effectively utilize limited GPU resources. Lastly, we are modifying the model's hyperparameters to find the most suitable conditions for Korean speech synthesis, aiming to improve the quality of the generated speech.

## ACKNOWLEDGMENT

The experiments in this paper were conducted using GPU resources provided by NIPA (National IT Industry Promotion Agency) and GPU resources purchased through the AI Robot Innovation Sharing Program of the National Research Foundation of Korea, supported by the Ministry of Education.